\documentclass[]{article}

\usepackage{layout}
\usepackage{amsmath}
\usepackage{textcomp}

\usepackage{graphics}
\usepackage{graphicx}

\newcommand{\be}{\begin{equation}}
\newcommand{\ee}[1]{\label{#1} \end{equation}}
\newcommand{\ba}{\begin{eqnarray}}
\newcommand{\ea}[1]{\label{#1} \end{eqnarray}}
\newcommand{\nl}{\nonumber \\}

\newcommand{\nbar}{\overline{n}}
\newcommand{\ebar}{\overline{e}}
\newcommand{\sbar}{\overline{s}}

\newcommand{\vs}{\vspace{3mm}}

\begin{document}

%% Place article title here:
%\title{Editorial policy and paper writing guide for \LaTeX ~users}
\title{{\bf Non-extensive equilibration in relativistic matter}}

%% Place for inserting article cathegory: Research Article, Rapid Communication, Communication or Review Article
%\articletype{Research Article}

\author{
     {T.~S.~Bir\'o and G.~Purcsel}\\
     {KFKI Research Institute for Particle and Nuclear Physics,}\\
     {P.O.Box 49, H-1525 Budapest, Hungary} 
}

\date{}

\maketitle

%% Please type your abstract here.
\abstract{
  We present a view of the non-extensive thermodynamics based on general composition rules.
  A formal logarithm maps these rules to the addition, which can be used to generate
  stationary distributions by standard techniques. We review the most commonly used
  rules and as an application we discuss the Tsallis-Pareto distribution of
  transverse momenta of energetic hadrons, which emerge from relativistic heavy-ion collisions.
}

\normalsize

\section{Introduction}

In non-extensive systems those thermodynamic variables, which usualy scale with system size
(with volume $V$ and particle number $N$), violate somehow this scaling. Generally
the entropy and energy of an $N$-particle system, $S_N$ and $E_N$,  are composed from 
individual quantities, $S_1$ and $E_1$, by counting for interaction and correlation
corrections. Considering pair interactions, mediated by the pair-potential  $v(r)$
(which depends on the relative coordinates $r=r_1-r_2$), and a particular form of
the two-particle phase space occupation, $\rho_{12}=f(p_1)f(p_2)g(r)$ with
the pair correlation function $g(r)$, approaching one for uncorrelated pairs, 
the following relations can easily be derived:
\ba
S_N / N = S_1 -   \nbar \int\!  g \ln g \, \, d^dr, \nl
E_N / N = E_1 +   \nbar \int\! g v \, \, d^dr,
\ea{SKETCH}
while the volume and the particle number are related via the mean density: 
$ V = N/\nbar = \int d^dr$.
In the above expression $S_1=2\sbar\int g(r) \, d^dr$ with $\sbar=-\int f(p) \ln f(p) \, d^dp$
and $E_1=2\ebar\int g(r) \, d^dr$ with $\ebar=\int f(p) K(p) \, d^dp$, where $K(p)$ denotes
the kinetic energy of a single particle.
In principle the correlation $g(r)$ can be obtained from the interaction $v(r)$ in a stationary state
(if exists), but this is a very complicated and difficult calculation for most of the known physical
systems. Non-extensivity occurs, whenever the specific ratios, like $E_N/N$ and $S_N/N$ diverge
in the $N\rightarrow\infty$ limit. One considers this limit at constant mean
number density, $\nbar$.

\vs
It is easy to construct examples for non-extensive energy at extensive entropy: in the crude approximation,
when $g(r)$ is either zero (up to a characteristic short range length) or one (towards infinity),
the entropy correction is zero and hence the large-$N$ entropy is extensive. However, with a
pair potential of the form $v(r)\sim r^{-\alpha}$, whenever $\alpha \le d$ with $d$ being the spatial
dimension ($d=3$ for isotropic systems), the correction to the energy becomes $N$-dependent
in the large-$N$ limit \cite{TsallisPrivCom}:
\be
 E_N / N \approx E_1 + {\rm const.} \, N^{1-\alpha/d}. 
\ee{NONEXTENER}
In systems with long range correlations (with scale independence, like in some random networks)
$g(r)$ may differ from one even at large distances. In such cases the entropy also may pick up
a non-extensive contribution. Another example may be given by confinement: the correlation
$g(r)$ approaches zero in this case for confined pairs  
at large relative distances, but the integral of $g\ln g$
may have a non-vanishing contribution which -- if of power-law type -- may diverge
with the total volume logarithmically.

\vs
The composition of small sytems into a big one (extensivity) and the composition of two
large systems (additivity) are related problems. Non-extensive systems are always non-additive
by using the original definitions for energy and entropy. In a quite broad class of cases,
however, another additive quantity may be constructed. The existence of such a quantity relies on
special properties (specifically on the associativity) of the composition rule.
The mapping of a non-additive quantity to an additive one, the formal logarithm, usually contains
parameters, which describe the degree of non-extensivity. This is the basis of the construction
of non-additive entropy (and energy) formulas.

\vs
This way a basic problem occurs for any non-extensive thermodynamics: how do large subsystems
equilibrate, whose thermal state is described not only by a temperature, $T$, but also by a
non-extensivity parameter, say $q$. In particular, for the Aczel-Daroczy-Chrvat-Tsallis
entropy formula, how does equilibration occur between different $(q,T)$ systems?
Given two preheated systems, does a common stationary distribution occur, will it be a Tsallis-Pareto
distribution, and is the temperature, defined by this equilibration process, universal
(absolute)? We attacked such questions in the framework of a particular parton cascade
model, using non-extensive energy composition rules in a Boltzmann equation type
simulation  \cite{BiroPurcselPLA} recently.

\vs
We note that the often cited entropy formula\cite{TsallisEntropyFormula}, 
\be
 S_T =  \frac{1}{1-q} \sum_i (w_i^q-w_i),
\ee{TSALLISENTROPY}
using normalized probabilities $\sum_i w_i=1$, 
follows the special composition rule
\be
 S_{12} = S_1 + S_2 + (1-q) S_1 S_2.
\ee{COMPOSETSALLIS}
For factorizing probabilities, $w_{ij}^{(12)}=w_i^{(1)}w_j^{(2)}$,
it can be mapped to an additive rule for $S_R=Y(S_T)$, given as
\be
 S_R = \frac{1}{a} \ln \left( 1 + a S_T \right).
\ee{SRENYI}
Here we used the parameter $a=(1-q)$. 
The result is the well-known R\'enyi entropy, 
\be
 S_R = \frac{1}{1-q} \ln \left( \sum_i w_i^q\right),
\ee{RENYIENTROPY}
which is additive but still contains
the extra parameter $q$. For this additive, and hence extensive, entropy formula the question
towards the two-parameter equilibration also holds \cite{Nauenberg}.

%%%%%%%%%%%%%%%%%%%%%%%%%%%% COMPOSITION RULES %%%%%%%%%%%%%%%%%

\section{Abstract composition rules generalize \newline non-extensivity}

The infinite repetition of an arbitrary pairwise, iterable
composition rule is an {\em associative} rule \cite{BiroComingSoon}.
It is a mathematical property that associative rules always possess
a strict monotonic function, called here the formal logarithm, in terms of which they can be 
expressed\cite{MATH_OF_FORMAL_LOG}.
We denote an abstract pairwise composition rule by the mapping 
$(x,y) \rightarrow h(x,y)$. 
The associativity of such a rule is expressed by 
\be
 h(h(x,y),z) \: = \: h(x,h(y,z))
\ee{ASSOC}
for $x,y$ and $z$ being real quantities. 
The general solution of the associativity equation (\ref{ASSOC}) is given by
\be
 h(x,y) = X^{-1}\left( X(x)+X(y)\right)
\ee{FORMLOG}
with  $X(x)$ being a strict monotonic function, the formal logarithm. 
It maps the arbitrary composition rule $h(x,y)$ to the addition by taking the 
$X$-function of eq.(\ref{FORMLOG}):
\be
 X(h(x,y)) = X(x) + X(y).
\ee{FORMADD}
Due to this construction there are generalized analogs to classical extensive (and additive) quantities;
they are formal logarithms. As a consequence stationary
distributions, in particular by solving generalized Boltzmann equations \cite{NEBE}, are 
proportional to the Gibbs exponentials of the formal logarithm,
\be
 f(x) = \frac{1}{Z} e^{-\beta X(x)}.
\ee{GIBBSFORM}
A general non-additive entropy formula can be derived based on the inverse of the
formal logarithm (inverting the $f \sim \exp \circ X$ function):
\be
 S = \int f X^{-1}(-\ln f).
\ee{ENTROPY_FORMULA}
The rule leading to the $q$-exponential distribution is given by
$h(x,y)=x+y+axy$ with the parameter $a$ proportional to $q-1$. In this case one obtains
the formal logarithm as being $ X(x) = \frac{1}{a} \ln(1+ax)$.
This formal logarithm leads to a stationary distribution with power-law tail
as the function composition $exp \circ X$ on the power $-\beta$:
\be
 f(E) = \frac{1}{Z} e^{- \frac{\beta}{a} \ln(1+a E)} = \frac{1}{Z} \left( 1+a E\right)^{-\beta/a}.
\ee{TSALLISEXP}
The corresponding non-additive entropy formula is constructed
as the expectation value of the inverse of this function,
$L^{-1} \circ \ln$ of $1/f$:
\be
 S = \int\! f \, \frac{e^{-a\ln(f)}-1}{a} \: = \: \frac{1}{a} \int  \, (f^{1-a}-f).
\ee{TSALLIS_ENTROPY}
The R\'enyi entropy is the formal logarithm of the Acz\'el-Dar\'oczy-Chrvat-Tsallis
entropy belonging to this composition rule.

\vs
Further examples for non-additive rules can be easily given. The power-rule,
$h(x,y)=(x^b+y^b)^{1/b}$, leads to a strecthed exponential, $f(x)\propto \exp(-\beta x^b)$,
in the stationary state. Kaniadakis \cite{KaniadakisRule} 
suggested a composition rule,
$h(x,y)= x\sqrt{1+\kappa^2y^2} + y\sqrt{1+\kappa^2x^2}$ with the corresponding formal 
logaritm being the inverse sine hyperbolic function, $X(x) = \frac{1}{\kappa} {\rm Ar sh} (\kappa x) $. 
The stationary distribution, 
$f(x) =  \frac{1}{Z} \left(\kappa x + \sqrt{1+\kappa^2x^2}\right)^{-\beta/\kappa} $,
develops a power-law tail for large $|x|$.
The corresponding entropy formula is the average of $X^{-1}\circ \ln$ 
of $1/f$ over the allowed phase space:
\be
 S_K = - \int \frac{f}{\kappa} \sinh(\kappa\ln f) = \int \frac{f^{1-\kappa}-f^{1+\kappa}}{2\kappa}.
\ee{KANI_ENT}
Regarding $\kappa x = p/mc$, the formal logarithm is proportional to the rapidity.
This would imply a stationary distribution like $exp(-|\eta|)$ with $\eta$ being the
rapidity.

\vs
Finally we note that the Tsallis rule $h(x,y)=x+y+axy$ is particular, being the most general 
symmetric second order formula satisfying $h(x,0)=x$.

\section{Equilibration of large subsystems}

Seeking for a canonical equilibrium state we have to maximize a total entropy given by
a general composition rule, $S(E_1,E_2)$, at the same time satisfying a constraint
which is in the general case also non-additive: $h(E_1,E_2)$ is constant.
For the moment we neglect the dependence on further thermodynamical variables;
usually the particle number $N$ and the volume $V$ is regarded to be proportional
and extensive.

In the traditional case both the entropy and the energy are combined
additively: $S(E_1,E_2)=S(E_1)+S(E_2)$ and $h(E_1,E_2)=E_1+E_2$. In the general case
by using corresponding formal logarithms the quantities $Y(S)$ and $X(E)$ have to be
considered as additive. Since for associative rules the formal logarithm is srict monotonic,
the maximum of the total entropy is achieved where $Y(S)$ has its extremum.
The general canonical principle is therefore given by
\be
 Y(S) - \beta X(E) = {\rm max.}
\ee{CANONICAL}
The parameter $\beta$ at this point is a Lagrange multiplier. Applying this for the equilibration
of two large subsystems, and assuming that the entropy of each systems depends only on its
own energy, one arrives at the equilibrium condition
\be
 \frac{Y'(S(E_1))}{X'(E_1)} \, S'(E_1) =
 \frac{Y'(S(E_2))}{X'(E_2)} \, S'(E_2) = \frac{1}{T}.
\ee{EQUIL_OF_TWO}
Comparing this with the general canonical form eq.(\ref{CANONICAL}) we obtain that $\beta=1/T$,
and $T$ is an absolute temperature in the clasiscal thermodynamical sense.
Its relation to the entropy, however, has been generalized.
In particular for an additive entropy, but non-additive energy composition rule,
one arrives at $1/T=S'(E)/X'(E)$. The relation of this quantity to the logarithmic
spectral slope, $1/T_{{\rm slope}}=-d \ln f/dE$ leads to a practical tool for the
analysis of particle spectra in experiments. For the Pareto-Tsallis disitribution
it is given by $T_{{\rm slope}}=T/X'(E)=T(1+aE)=T+(q-1)E$. The naive effort to extract a
temperature from energy spectra of particles, as it is a widespread praxis in relativistic
heavy ion studies, only works if $q=1$, i.e. for spectra exponential in the particle energy.
Otherwise an energy dependent slope, and a curved spectrum in the logarithmic plot
has to be interpreted.

\section{Spectral temperatures in relativistic heavy ion collisions}

It is helpful to describe shortly, how a temperature can be conjectured from
observations on particle spectra produced in relativistic heavy ion collisions.
The detected particles have relativistic velocities and different masses.

\vs
One intriguing way is to look at the transverse momentum, $p_T$-, spectra around
mid-rapidity (in the center of mass system for equal colliding heavy ions).
The different identified hadrons, mostly pions, kaons, protons and antiprotons,
have to show that their abundance in the momentum space depends on their energy;
this phenomenon at zero rapidity is the so-called $m_T$-scaling.
The transverse mass is given as $m_T=\sqrt{m^2+p_T^2}$, at strictly zero
rapidity this is the total relativistic energy.

\begin{figure}
\centerline{\includegraphics[width=0.45\textwidth,angle=-90]{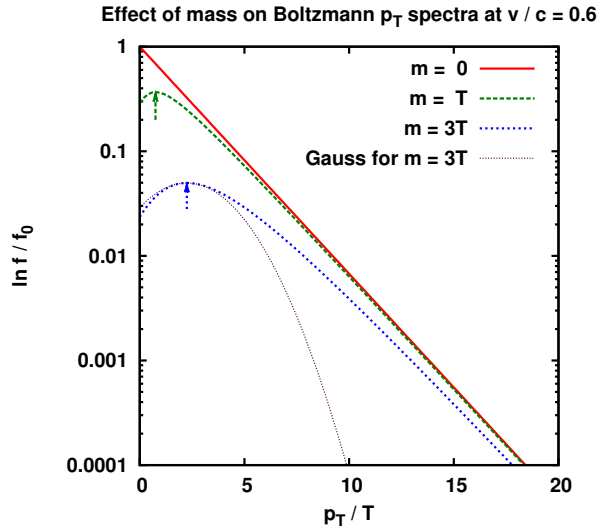}}
\caption{\label{Fig1}
  General shape of $p_T$ spectra for massive particles in the presence of a bulk transverse flow
  with radial velocity component $v_T=0.6$. The Gaussian approximation to the maximum reveals
  both a Lorentz-enhanced mass, $m^*=m\gamma_T$, and a Lorentz-enhanced temperature parameter,
  $T\gamma_T$.
}
\end{figure}

\vs
The analysis is made a little more involved by the fact that the source emitting the
detected hadrons is not at rest. The most prominent feature is a transverse flow,
with relativistic velocity, $v_T$ (and a corresponding Lorentz factor $\gamma_T=1/\sqrt{1-v_T^2}$
in units where $c=1$). The relativistic energy of a particle in the frame of the emitting
source cell is given by the J\"uttner variable: 
\be
E=u_{\mu}p^{\mu}=\gamma_T m_T \cosh(y-\eta)-\gamma_Tv_Tp_T\cos(\varphi-\Phi).
\ee{JUTTNER}
Here the four-velocity of the source and the actual four-momentum of the particle
are parametrized by rapidity and angle variables:
\ba
 u_{\mu} &=& (\gamma_T \cosh \eta, \gamma_T \sinh \eta, \gamma_Tv_T \cos \Phi,  \gamma_Tv_T \sin \Phi),
\nl
 p_{\mu} &=& (m_T \cosh y, m_T \sinh y, p_T \cos \varphi, p_T \sin \varphi). 
\ea{DEF_FOUR_VECTORS}
We consider a thermal model for the particle spectra; then the yield is supposed to
depend on the J\"uttner variable $E$ given by eq.(\ref{JUTTNER}).
Assuming a general distribution $f(E) \sim exp(-X(E)/T)$, which is monotonic decreasing, 
one finds its maximum at the minimum of $E$. This variable is minimal at the rapidity
$y_{{\rm min}}=\eta$, and angle $\varphi_{{\rm min}}=\Phi$, giving 
\be
  E_{{\rm min}}=\gamma_Tm_T-\gamma_Tv_Tp_T. 
\ee{MIN_TRV_ENERGY}
This Lorentz-boosted transverse energy reaches its minimum at the 
transverse momentum value \hbox{$p_{T,{\rm min}}=m\gamma_Tv_T$,}
leading to \hbox{$m_{T,{\rm min}}=m\gamma_T$} and \hbox{$E_{{\rm min}}=m$.} 
The expansion around this minimum in the $p_T$-distribution is an effective Gaussian:
\be
 e^{-(E-m)/T} \approx \exp\left({-\frac{(p_T-m\gamma_Tv_T)^2}{2m\gamma_T \, T\gamma_T}}\right).
\ee{EFF_GAUSS}
Such spectra are plotted in Fig.\ref{Fig1} for a typical transverse flow of $v_T=0.6$
for massless and massive particles with masses of $m=T$ and $m=3T$. 
The curves show typical shapes for light meson and heavy baryon spectra occuring in
relativistic heavy ion collisions.

%%%%%%%%%%%%%%%%%%%%%%%%%%%%%%%%
%%%%%%%%%%%%%%%%%%%%%%%%%%%%%%%%%%%%%%%%%%%%%%%%%%%%%%%%%%%%%%%%%%%%%%%%%%%

In fact, according to experimental findings at RHIC the observed particle spectra have to be
corrected for a transverse flow in order to reach $m_T$-scaling. On the other hand
the formula near the maximum, eq.(\ref{EFF_GAUSS}), may shed some light to the
classical Einstein-Ott-Planck discussion about the temperature of relativistically
moving bodies from an unexpected corner of modern experimental observations. 
Namely by using the Gaussian approximation both the particle mass
and the effective spectral temperature gain a Lorentz factor, $\gamma_T$.

\section{Non-extensivity in quark matter and in hadron matter}

We conjecture that the power-law tails observed in hadronic spectra may stem from the
non-extensivity of the preformed quark matter, which hadronizes rapidly.
We make a connection between quark and hadron spectra by the quark coalescence model.
A coalescence of two quarks (actually a quark and an antiquark) into a meson produces
a yield proportional to the following quantity:
\be
 F(\vec{p}) = \int f\left(E(\vec{P}/2+\vec{q})\right) f\left(E(\vec{P}/2-\vec{q})\right) C(\vec{q}) \, d^3q. 
\ee{COALESCENCE}
Here we integrate over the relative momentum of the quarks with a coalescence
factor, $C(\vec{q})$, for which a simple model has been utilized \cite{JPG.COAL}.
For common momenta much larger than the relative one $|\vec{P}|\gg |\vec{q}|$
(otherwise the quarks do not coalesce!) on gets
\be
 F(\vec{P}) \approx \,  f^2\left(E(\vec{P}/2)\right) \int  C(\vec{q}) \, d^3q. 
\ee{LARGE_MOM_COAL}
In particular light hadrons made from massless quarks follow the quark-scaling rule:
\be
 f_{{\rm hadron}}(E) \propto f^n(E/n).
\ee{QUARK_RULE}
As a consequence particular properties of the non-extensive thermal model between quark and
hadron matter also scale:
$T_{{\rm mesons}} = T_{{\rm baryons}} = T_{{\rm quarks}}$ for the temperature, while 
  $q_{{\rm mesons}}-1 = (q_{{\rm quarks}}-1)/2$ for mesons and 
  $q_{{\rm baryons}}-1 = (q_{{\rm quarks}}-1)/3$ for baryons. 
Experimentally these relations are still to be checked.
These predictions of the non-extensive phenomenology 
meet the curves from pQCD calculations smoothly, with the following
surmised properties of quark matter at RHIC: $T=140\ldots 180$ MeV, $q=1.22$, $v_T=0.6$. 
\cite{SQM2007}.

\vspace{3mm}
Summarizing we have shown that non-extensive behavior can be mapped to additive properties
of a formal logarithm of the original quantity in the general case of associative composition rules.
That such rules necessarily arise in the thermodynamical limit is demonstrated in 
Ref.\cite{BiroComingSoon}.
Using formal logarithms all the classical concepts and techniques can be applied to describe
thermal equilibrium or to generate distributions accordingly. The thermal equilibration of
large subsystems are subject to straightforward generalizations of the familiar rules. 
As an exaple we discussed certain
particle spectra arising in relativistic heavy ion collisions from a thermal and non-extensive
quark matter. These results qualitatively agree with experimental findings.

\vspace{3mm}
{\bf Acknowledgement}

This work has been supported by the Hungarian National Science Fund, OTKA
(K49466, K68108). Discussions with
C.~Tsallis, G.~Kaniadakis, P.~H\"anggi, P.~V\'an, K.~\"Urm\"ossy, G.~G.~Barnaf\"oldi
are gratefully acknowledged.

%%%%%%%%%%%%%%%%%%%%%%%%%%%%%%%%%%% BIBLIOGRAPHY %%%%%%%%%%%%%%%%%%%%%%%%%%%%%%%%

\end{document}